\documentclass{vietnam}

\def\be{\begin{equation}}
\def\ee{\end{equation}}
\def\bea{\begin{eqnarray}}
\def\eea{\end{eqnarray}}

\newcommand{\Photo}{\includegraphics[height=0mm]{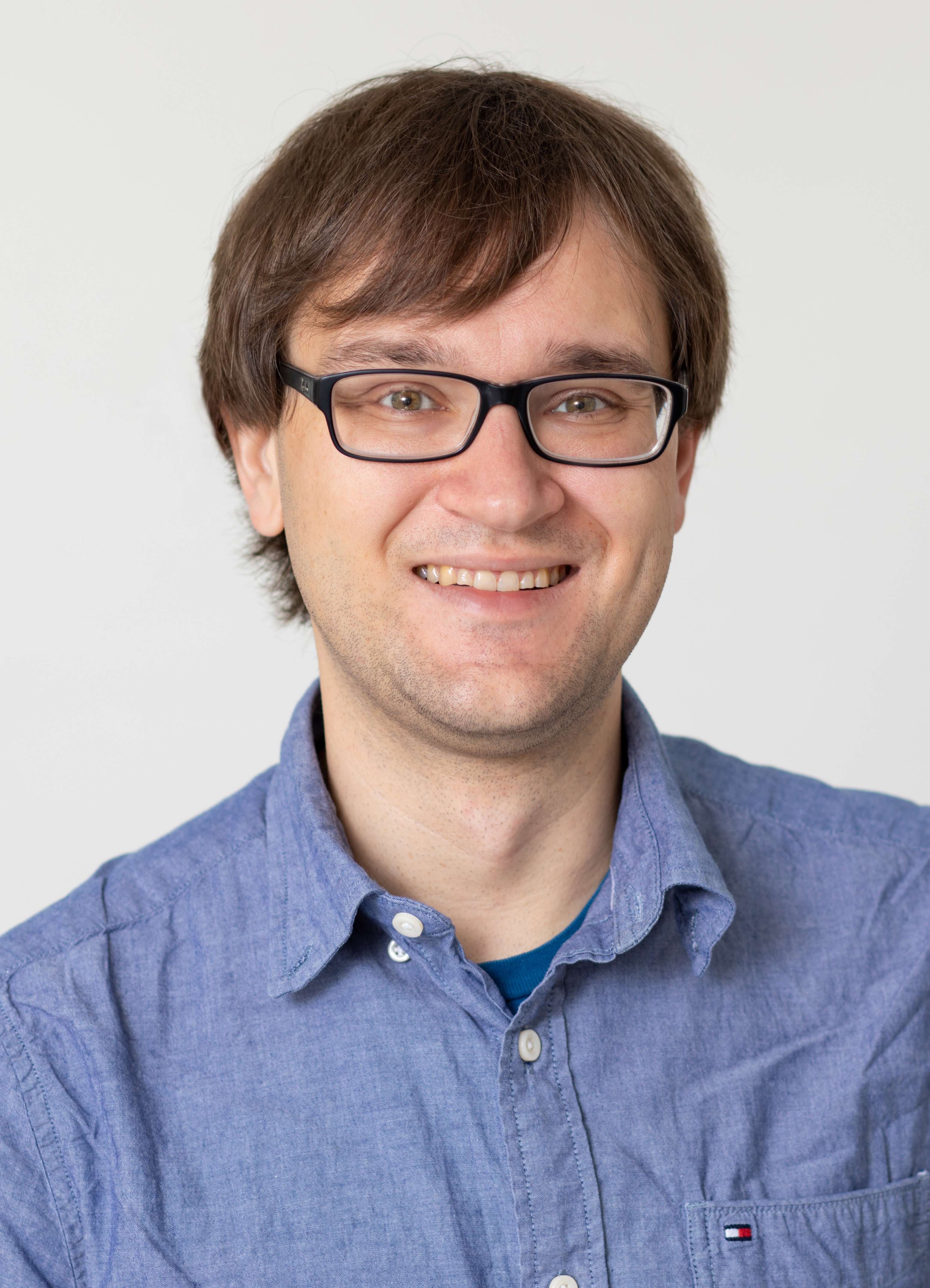}}

\begin{document}
\vspace*{4cm}
\title{{\Large Neutrino Physics at the LHC: Status and Prospects} \\ \medskip 
\textnormal {A proceeding for Rencontres du Vietnam / PASCOS~2024
}}
\author{Felix Kling}
\address{Deutsches Elektronen-Synchrotron DESY, \\Notkestr.~85, 22607 Hamburg, Germany \vspace{2mm}} 

\maketitle\abstracts{The LHC is not only the most powerful collider built to date but also the source of an intense beam of the most energetic neutrinos ever produced by humankind. After nearly 15 years of LHC operation, these neutrinos have been observed for the first time by the FASER and SND@LHC experiments. This breakthrough marks the dawn of a new field: collider neutrino physics. Further neutrino measurements at the LHC will offer novel opportunities to advance neutrino physics, constrain the strong interaction in uncharted kinematic regimes, provide critical input for addressing outstanding questions in astroparticle physics, and search for phenomena predicted by scenarios of physics beyond the Standard Model. This proceeding reviews the existing and proposed neutrino detectors, presents their first results, and summarizes their physics potential.}

\section{Introduction}

The Large Hadron Collider (LHC) is the most powerful particle collider built thus far, with a collision energy of 13.6~TeV. Its primary focus is the study of heavy Standard Model (SM) particles, such as the Higgs boson, and the search for other heavy particles predicted by models of physics beyond the Standard Model (BSM), such as models of supersymmetry. The multi-purpose detectors, such as ATLAS and CMS, were optimized to catch such signals.

\begin{figure}
\centering
\includegraphics[width=0.8\linewidth]{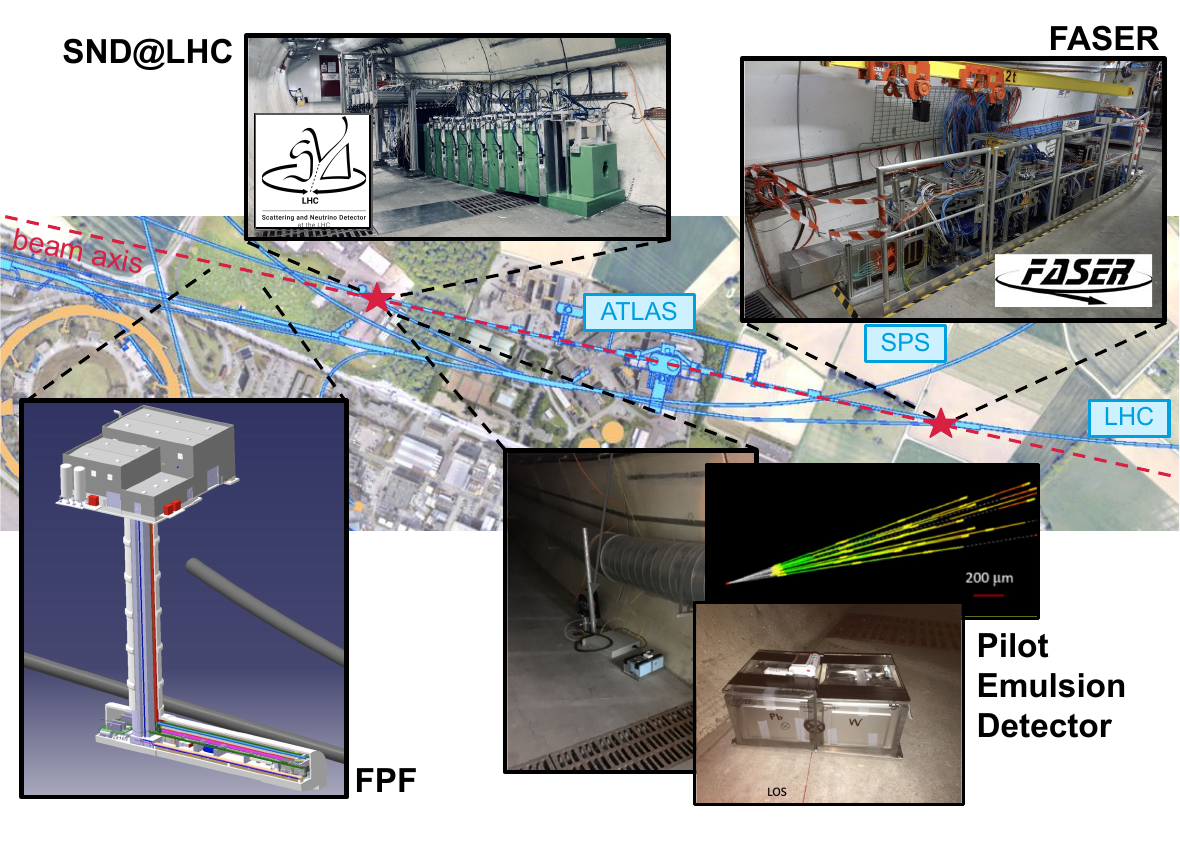}
\vspace*{-5mm}
\caption{\textbf{Location of the Collider Neutrino Experiments.} Primary collisions at the ATLAS IP create a strongly collimated neutrino beam along the beam axis (red dashed). This beam axis is accessible in two side tunnels, connecting the SPS and LHC rings, about 480~m downstream of ATLAS (red stars). Pictures of the pilot emulsion detector, the FASER experiment as well as the SND@LHC experiment in these locations are shown. The proposed FPF would be located 620~m west of ATLAS.}
\label{fig:location}
\end{figure}

Already in the 1980s, it was noticed that the LHC also produces a large number of neutrinos~\cite{DeRujula:1984ns}. Indeed, at each interaction point (IP), the LHC generates an intense and strongly collimated beam of high-energy neutrinos along the beam collision axis. While the LHC tunnel eventually curves away, the neutrinos will continue to propagate straight along the beam axis. As illustrated in Fig.~\ref{fig:location}, roughly 480 m downstream from the ATLAS IP, this beam axis intersects with the TI12/TI18 tunnels which used to host the injectors of the Large Electron-Positron Collider (LEP) but remained unused during the LHC era. Therefore, this location provides a rare opportunity to access the beam collision axis and exploit the neutrino beam.

This opportunity was first realized by the FASER collaboration which proposed to install a detector to search for light long-lived particles in 2018~\cite{Feng:2017uoz,FASER:2018ceo,FASER:2018bac,FASER:2018eoc,FASER:2019aik}. In the summer of that year, a small emulsion pilot detector with a target mass of 11~kg was placed in the TI18 tunnel for just 4 weeks --- see photo in Fig.~\ref{fig:location}. Using the collected data, the FASER collaboration announced the first detection of collider neutrino candidates in 2021~\cite{FASER:2021mtu}. While the reported significance of 2.7$\sigma$ was not enough to claim an observation of neutrinos at the LHC, this measurement illustrates how detectors at this location can complement the large multi-purpose detectors at the LHC~\cite{Foldenauer:2021gkm}. 

During the Long Shutdown~2 (2019 - 2022), two detectors were installed to take advantage of this opportunity: FASER~\cite{FASER:2019dxq} and SND@LHC~\cite{SHiP:2020sos}. These experiments have been taking data since the summer of 2022 and have since then reported the first observation of collider neutrinos~\cite{FASER:2023zcr,SNDLHC:2023pun}. They will continue to operate until the end of Run~3 of the LHC in 2026, and upgrades are envisioned for operation in Run~4 starting in 2030. A continuation and expansion of this experimental program during the high-luminosity LHC (HL-LHC) era with significantly larger detectors has been proposed in the context of the Forward Physics Facility (FPF)~\cite{Anchordoqui:2021ghd,Feng:2022inv,Adhikary:2024nlv}. Pictures of both detectors as well as the design of the FPF are also shown in Fig.~\ref{fig:location}.

\section{Experimental Program}

\begin{figure}
\centering
\includegraphics[width=0.89\linewidth]{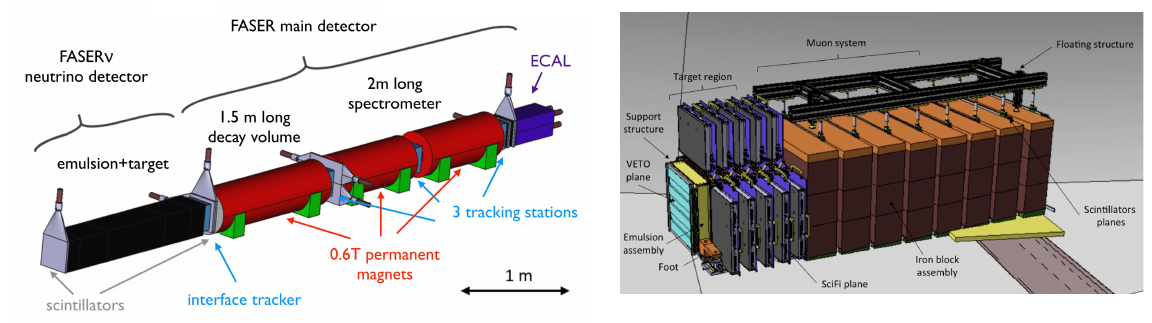}
\vspace*{-3mm}
\caption{\textbf{Design of the FASER and SND@LHC.} The left figure shows the FASER experiment, consisting of the dedicated FASER$\nu$ emulsion detector at its front followed by electronic components: a magnetized spectrometer and calorimeter. The right panel shows the SND@LHC detector. The target region at its front consists of an emulsion detector interleaved with SciFi tracking planes, which are followed by a hadronic calorimeter.}
\label{fig:experiments}
\end{figure}

\noindent \textbf{FASER detector:} The FASER experiment is located in TI12 where a concrete trench has been excavated to align the detector precisely with the beam axis~\cite{FASER:2022hcn}, covering the pseudorapidity region $\eta > 9$. A schematic layout of the experiment is shown in the left panel of Fig.~\ref{fig:experiments}. At the front of the experiment is the FASER$\nu$ neutrino detector~\cite{FASER:2020gpr}. It consists of a 1.1~ton tungsten target interleaved with emulsion films, which enable precise reconstruction of neutrino interactions. On its upstream end, a front veto consisting of two scintillator layers detects incoming charged particles, while downstream, an interface tracker connects the emulsion detector with the electronic components of the FASER main detector.  The FASER main detector consists of a veto station, followed by and a 1.5~m long magnetized decay volume and tracking spectrometer~\cite{FASER:Tracker}. Further downstream, the pre-shower station provides particle identification capabilities while the electromagnetic calorimeter, composed of four spare modules from the LHCb experiment, measures the energy of electromagnetic showers. \smallskip

\noindent \textbf{SND@LHC} The SND@LHC experiment is placed in the TI18 tunnel, on the opposite site of ATLAS~\cite{SNDLHC:2022ihg}. Unlike FASER, it is placed off-axis and covers the pseudorapidity region of $7.2< \eta<8.4$. A layout of the detector is shown in the right panel of Fig.~\ref{fig:experiments}. The detector's target region is composed of a hybrid system based on an 830~kg target made of tungsten plates, interleaved with emulsion and electronic trackers, which also act as an electromagnetic calorimeter. It is preceded by a veto scintillator and followed by a hadronic calorimeter that also acts as a muon identification system. \smallskip

\noindent \textbf{The Forward Physics Facility} While FASER and SND@LHC have successfully begun operations in the former service tunnels TI12 and TI18, it is important to note that these locations were not originally intended to host experiments or their associated infrastructure. Furthermore, their limited size prevents any expansion, making it impossible to accommodate larger detectors or additional experiments. To overcome these constraints, the concept of the FPF has been proposed as a dedicated space for a suite of experiments during the HL-LHC era~\cite{Anchordoqui:2021ghd}. Extensive site studies identified a suitable location approximately 620~m west of the ATLAS interaction point (IP), where a cavern roughly 65~m long and 5~m wide, aligned with the beam collision axis, would be excavated~\cite{Feng:2022inv,Adhikary:2024nlv}. The FPF is designed to host several next-generation experiments, including the large-volume long-lived particle detector FASER2, the emulsion-based neutrino detector FASER$\nu$2, the dedicated milli-charged particle detector FORMOSA~\cite{Foroughi-Abari:2020qar}, and the liquid argon neutrino detector FLARE. Notably, both neutrino detectors will allow target masses of 10~tons and more, and therefore significantly increase the number of neutrinos events than can be record with them in comparison to the 1~ton targets of the currently operating detectors. In addition, the FASER2 experiment will also act as a muon spectrometer for the neutrino experiments. 

\section{First Results}

\begin{figure}
\centering
\includegraphics[width=0.99\linewidth]{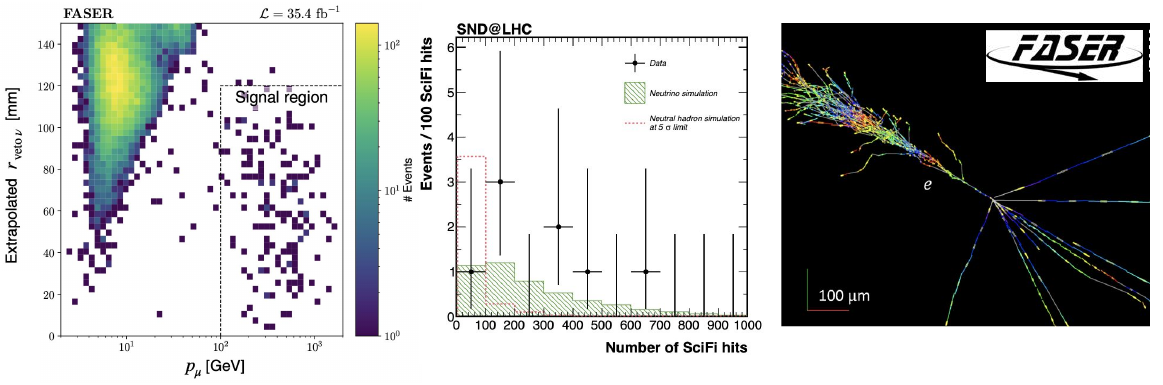}
\vspace*{-3mm}
\caption{\textbf{Observation of Collider Neutrinos.} The left and central panels shows the distribution of events selected by the neutrino observation analysis performed by FASER and SND@LHC, respectively. The right panel shows an electron neutrino candidate observed by FASER$\nu$. }
\label{fig:observation}
\end{figure}

\noindent \textbf{First Observation of Collider Neutrinos by FASER:} In March 2022, the FASER collaboration announced the first observation of neutrinos at the LHC~\cite{FASER:2023zcr} using the data set collected in 2022 with about 36~fb$^{-1}$. Since the processing of the emulsion detector data takes time, this analysis used only the electronic detector components~\cite{Arakawa:2022rmp} to search for charged current $\nu_\mu$ interactions via the appearance of an energetic muon. Events were selected for which the scintillators upstream of the FASER$\nu$ target were not activated and exactly one track emerged and passed through the spectrometer. Muon neutrino candidate events were required to have a momentum $p_\mu$ of at least 100~GeV and an extrapolated radial distance from the beam center at the front veto $r_{veto}$ of less than 12~cm. Backgrounds were investigated and found to primarily originate from low momentum muons entering close to the edge of the detector  and neutral hadrons interacting in the target. A total background of $0.08 \pm 1.83$ was estimated. After unblinding, a total of 153 events were observed in the signal region, corresponding to a significance of more than 15$\sigma$. Events with at least one track and no activity in the front veto are shown in the left panel of Fig.~\ref{fig:observation} in terms of $r_{veto}$ and $p_\mu$. One can observe a clear separation between the population of background events, consisting of low momentum muons entering from the side, and neutrino signal events, with high momentum in the central detector region.  \smallskip

\noindent \textbf{Observation by SND@LHC:} The observation of neutrinos at the LHC was confirmed by an additional search performed by the SND@LHC collaboration which was released in May 2025~\cite{SNDLHC:2023pun}. Similar to FASER, this analysis identified muon neutrino interactions via the appearance of a muon using only data from the electronic detector components collected in 2022. The analysis required no activity in the upstream veto system, a muon track in the muon system and a shower in the target area as observed in the SciFi tracker and hadronic calorimeter. A total of 0.076 events were expected to pass all selection cuts and 8 candidate events were identified in the data, corresponding to a significance of 7$\sigma$. The central panel of Fig.~\ref{fig:observation} shows the number of SciFi hits for selected events in data compared to predictions from simulation for signal and background. \smallskip

\noindent \textbf{First Observation of Electron Neutrinos by FASER$\nu$:} In March 2024, the FASER collaboration released the first results using the FASER$\nu$ emulsion detector~\cite{FASER:2024hoe}. This analysis used a small subset of the collected emulsion detector data, corresponding to a target mass of 128.8~kg and a luminosity of 9.5~fb$^{-1}$. This only corresponds to about 2\% of the data collected in 2022 and 2023. After being removed from the FASER$\nu$ detector, the emulsion films were developed, scanned, and digitized. Tracks are first reconstructed in each film and then aligned across films. Charged current neutrino interactions produce a neutral vertex signature, defined as a vertex with no incoming and at least five outgoing charged particle tracks, associated with an energetic electron or muon candidate. Electron candidates are identified using a high density of track segments produced in the electromagnetic shower, and their energy is estimated from the segment multiplicity. Muon candidates are selected as long tracks penetrating more than 100 tungsten plates, and their momentum is estimated using multiple Coulomb scattering. A possible background to neutrino signals arises from neutral hadron interactions, which are, however, typically lower in energy and are not associated with a high-energy lepton. It was found that this background can be reduced to negligible levels by requiring a lepton with reconstructed energy of over 200~GeV, a back-to-back topology of the lepton and hadron system, and additional topological cuts.  A total background, mainly coming from neutral current interactions, of 0.025 and 0.22 events was estimated for the $\nu_e$ and $\nu_\mu$ charged current interaction channels, respectively. Four data events were selected by the $\nu_e$ selection, and eight by the $\nu_\mu$ selection, in agreement with expectations. The statistical significance of this observation is 5.2$\sigma$ for $\nu_e$ and 5.7$\sigma$ for $\nu_\mu$. This measurement therefore constitutes the first observation of electron neutrino interactions at a collider. An example $\nu_e$ event is shown in the right panel of Fig.~\ref{fig:observation}. These results were also used to perform the first measurement of the neutrino interaction cross section at TeV energies.

\section{Neutrino Fluxes as Novel Probe of Forward Particle Production}

\begin{table}
\setlength{\tabcolsep}{5.0pt}
  \caption{Expected rate of charged current neutrino interactions in the LHC neutrino detectors.}
  \centering
  \begin{tabular}{c|c|c|c||c|c|c}
  \hline\hline
  \multicolumn{4}{c||}{Detector} & 
  \multicolumn{3}{c}{CC Interactions} \\
  \hline
  Name & Mass & Luminosity & Rapidity 
  & $\nu_e\!\!+\!\bar{\nu}_e$ 
  & $\nu_\mu\!\!+\!\bar{\nu}_\mu$
  & $\nu_\tau\!\!+\!\bar{\nu}_\tau$
  \\
  \hline\hline
  SND@LHC at Run~3
  & 0.8~t & 250~fb$^{-1}$ & $7.2<\eta<8.4$
  & 200 & 1.2k & 11 \\
  FASER$\nu$ at Run~3
  & 1.1~t & 250~fb$^{-1}$ & $\eta>8.8$
  & 1.6k & 8.5k & 32 \\
  FASER$\nu$ at HL-LHC
  & 1.1~t & 3~ab$^{-1}$ & $\eta>8.8$
  & 19k & 102k & 360 \\
  \hline
  FASER$\nu$2 at FPF
  & 20~t & 3~ab$^{-1}$ & $\eta>8.5$
  & 280k & 1.1M & 9.4k \\
  FLArE at FPF
  & 10~t & 3~ab$^{-1}$ & $\eta>7.5$
  & 68k & 250k & 3k \\
  \hline\hline
  \end{tabular}
  \label{tab:number}
\end{table} 

\begin{figure}[tbp]
\centering
\includegraphics[width=0.32\textwidth]{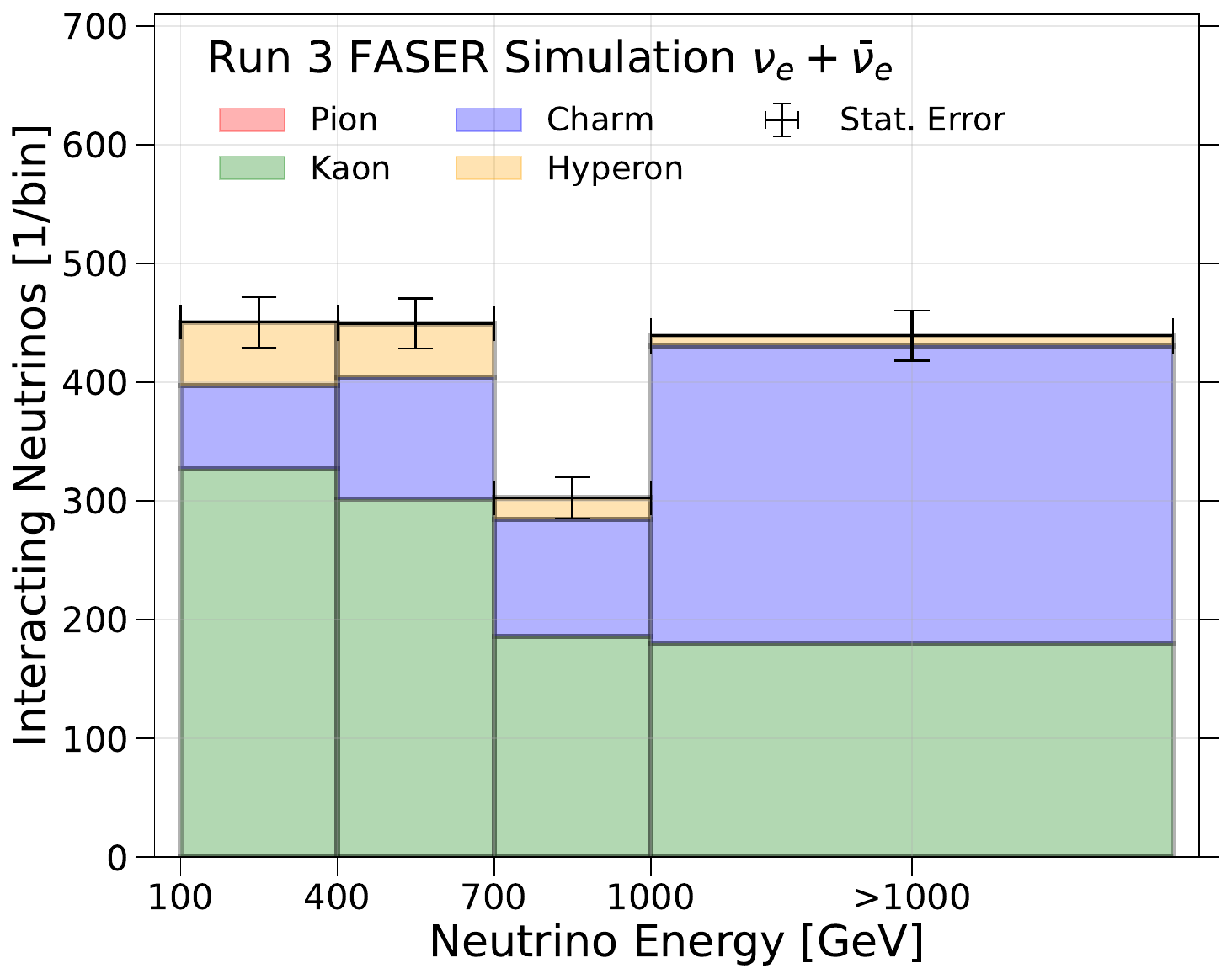}
\includegraphics[width=0.32\textwidth]{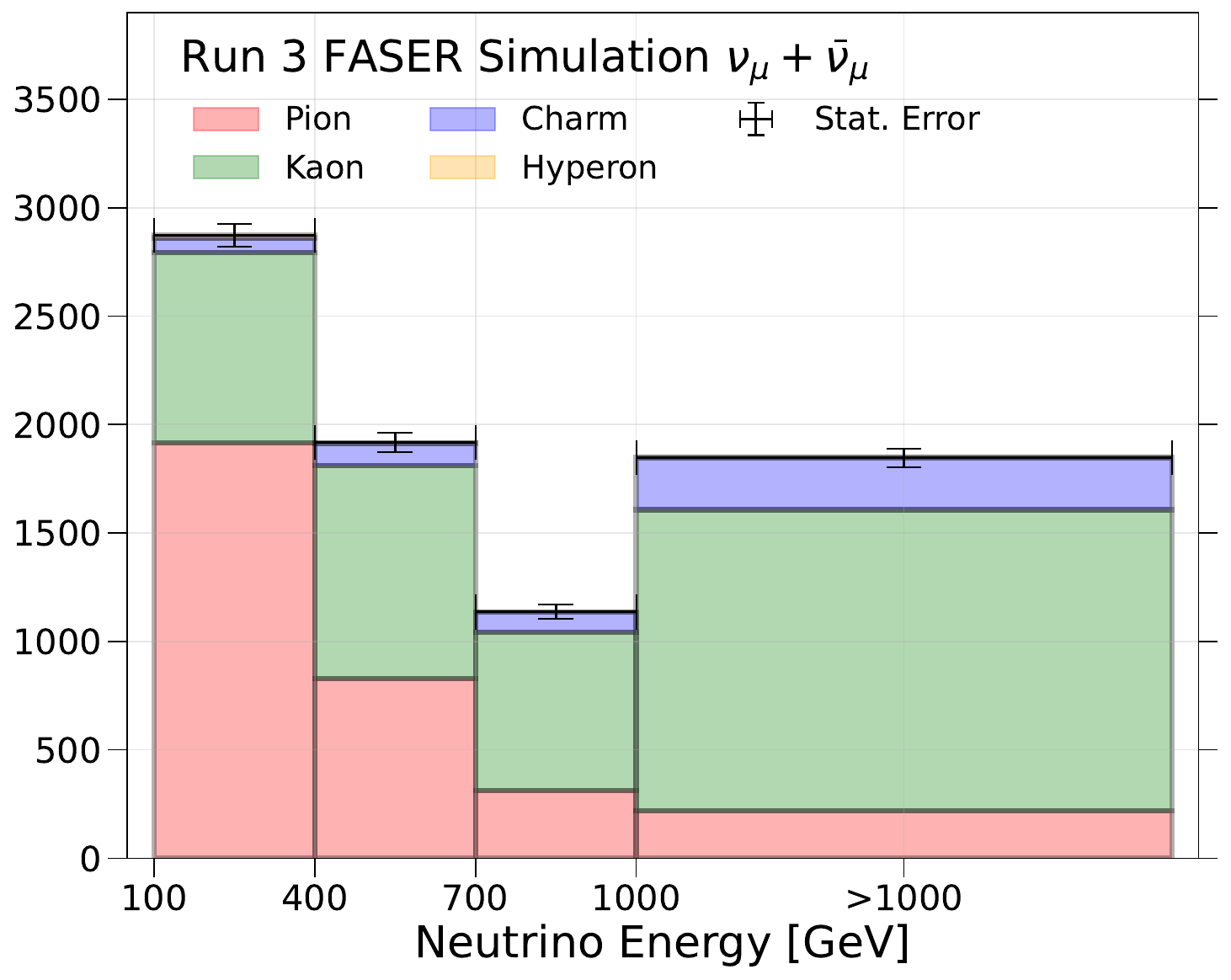}
\includegraphics[width=0.32\textwidth]{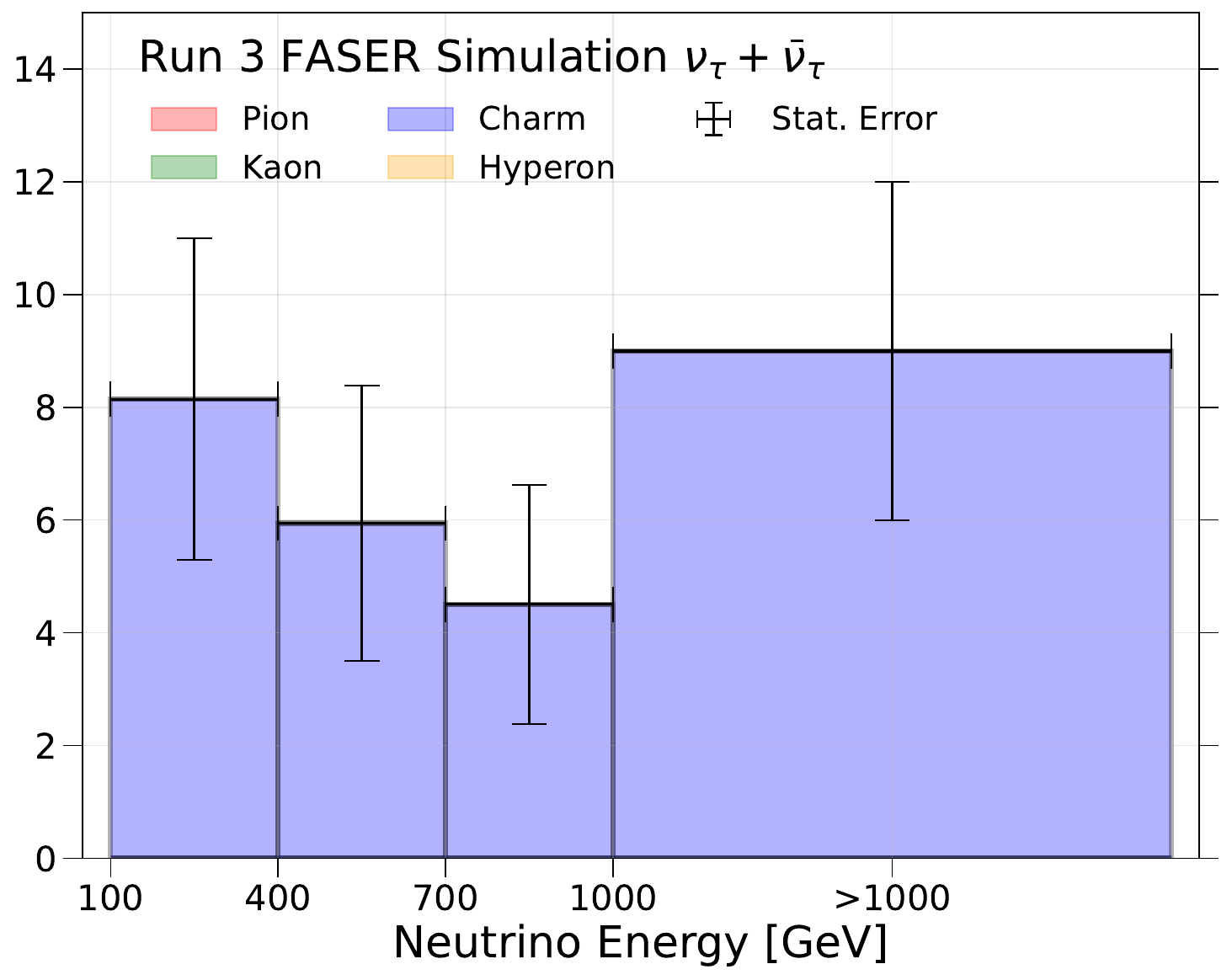}
\vspace*{-3mm}
\caption{\textbf{Origin of Neutrinos.} The panels show the binned energy spectra of interacting $\nu_e$ (left), $\nu_\mu$ (center), and $\nu_\tau$ (right) in FASER$\nu$ at LHC Run~3. The neutrinos are separated by their parent hadrons: pions (red), kaons (green), charm hadrons (blue), and hyperons (yellow).}
\label{fig:origin}
\end{figure} 

The first observation of neutrinos at the LHC marks the dawn of collider neutrino physics\cite{Worcester:2023njy}. As shown in Tab~\ref{tab:number}, the number of neutrinos observed with existing and proposed LHC detectors is expected to grow rapidly~\cite{Kling:2021gos,Buonocore:2023kna,FASER:2024ykc}. While only a few hundred neutrinos have been detected so far, current experiments are projected to record around ten thousand neutrinos by the end of Run~3 in 2026. In the HL-LHC era, detectors of similar size could see about hundred thousand neutrino interactions, while larger detectors proposed for the FPF could observe over a million collider neutrinos. As shown in Fig.~\ref{fig:origin}, collider neutrinos mainly originate from the decay of charged pions, kaons, and charm mesons. The forward production of these particles at LHC energies has not been measured before. Neutrino flux measurements thus provide a novel way to probe and constrain forward hadron production, offering unique insights into the strong interaction in previously inaccessible kinematic regions. \smallskip

\noindent \textbf{Hadronic Interaction Models and Astroparticle Physics:} Forward light-hadron production falls outside the validity region of perturbative QCD. Instead, their production is typically described using phenomenological hadronic interaction models that were tuned to the available data, as for example recently done in the case of Pythia~\cite{Fieg:2023kld}. Previous forward particle measurements at the LHC were performed by LHCf and are restricted to neutral pions and neutrons. Collider neutrino experiments will add complementary data on charged pions and kaons.

Improving forward particle production models is essential for astroparticle physics, where they are used to describe the production of high-energy particles in extreme astrophysical environments and the interactions of cosmic rays with Earth’s atmosphere. However, cosmic ray experiments have reported a significant mismatch between the observed number of muons in high-energy cosmic ray air showers and predictions from hadronic interaction models~\cite{Soldin:2021wyv}. Known as the \textit{muon puzzle}, this discrepancy currently prevents an accurate determination of the cosmic ray composition and origin. Decades of studies suggest this mismatch likely arises from a mismodeling of soft QCD effects in forward particle production at center-of-mass energies above the TeV scale~\cite{Albrecht:2021cxw}. Additional research indicates the problem could be resolved by an increased rate of forward strangeness production, leading to increased neutrino fluxes from kaon decays that could be tested at collider neutrino experiments~\cite{Anchordoqui:2022fpn,Sciutto:2023zuz,Kling:2023tgr}. \smallskip

\noindent \textbf{Forward Charm Production and QCD:} In contrast to light hadrons, forward charm production can be modeled using perturbative QCD methods. A variety of calculations using both the collinear factorization~\cite{Bai:2020ukz,Buonocore:2023kna} and k$_T$-factorization~\cite{Maciula:2022lzk,Bhattacharya:2023zei} frameworks have been presented in recent years. Although all estimates were guided by LHCb data, these resulting neutrino flux predictions differ by more than one order of magnitude. This illustrates how collider neutrino measurements provide novel and complementary information to test theory predictions.  

Charm quarks are predominantly produced via gluon fusion ($gg \to c\bar{c}$), where one gluon carries a large momentum fraction ($x \sim 1$), and the other a very small fraction ($x \sim 10^{-7}$). The high-$x$ region is sensitive to intrinsic charm~\cite{Maciula:2022lzk} and constrain high-$x$ PDFs, while low-$x$ region is sensitive to the gluon PDF in regions far beyond the reach of current experiments, enabling tests of QCD phenomena such as BFKL dynamics and gluon saturation~\cite{Bhattacharya:2023zei}. Improved understanding of forward charm production also benefits astroparticle physics, particularly by reducing uncertainties in the prompt atmospheric neutrino flux. This flux, arising from charmed hadron decays in cosmic ray collisions with the atmosphere, becomes the dominant background for astrophysical neutrino searches at energies above a few 100~TeV. Current uncertainties limit sensitivity in neutrino telescopes like IceCube, but forward charm measurements at the LHC can refine these predictions, enhancing astrophysical neutrino studies and multi-messenger astronomy~\cite{Adhikary:2024nlv,Bai:2022xad}.

\section{Collider Neutrino Experiments as Neutrino-Ion Collider}

\begin{figure}[tbp]
\centering
\includegraphics[width=0.95\textwidth]{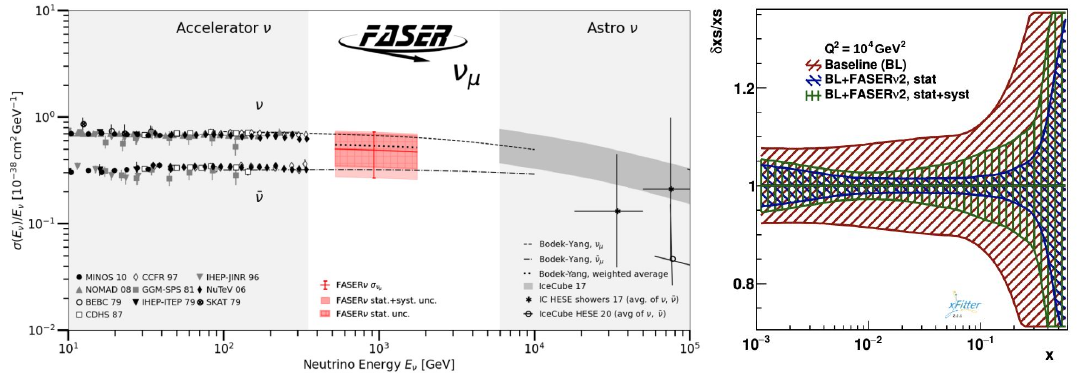}
\vspace*{-3mm}
\caption{\textbf{Neutrino Interactions.} The left panels shows the muon neutrino interaction cross section as measured by FASER$\nu$. The right panel shows the projected sensitivity to constrain the strange PDF via neutrino scattering in FASER$\nu$2.}
\label{fig:interaction}
\end{figure} 

The LHC neutrino beam's energy surpasses that of all artificial neutrino sources and is succeeded only by atmospheric and astrophysical neutrinos. This enables LHC neutrino experiments to measure neutrino interaction cross sections at previously inaccessible TeV energies. A first measurement has already been performed by FASER$\nu$~\cite{FASER:2024hoe} and are shown in the left panel of Fig.~\ref{fig:interaction}. Future measurements, performed for all three neutrino flavors, offer a unique opportunity to test lepton universality in neutrino scattering~\cite{Falkowski:2021bkq}. Additionally, the ability of emulsion detectors to observe tau neutrinos allows for precision studies of tau neutrino properties and the first separate observation of tau neutrinos and tau antineutrinos.

Forward neutrino experiments also establish the LHC as a \textit{neutrino-ion collider}, probing charge current deep-inelastic scattering (DIS) at center-of-mass energies between 10 and 50 GeV. This complements the planned electron-ion collider, which will focus on neutral current DIS in a similar energy range. The large neutrino sample enables differential cross-section measurements to constrain proton and nuclear parton distribution functions (PDFs). As shown in the right panel of Fig.~\ref{fig:interaction}, high-statistics neutrino scattering data could significantly reduce PDF uncertainties~\cite{Cruz-Martinez:2023sdv}. This improvement will benefit key measurements at the LHC central detectors, such as Higgs or weak boson production, and help disentangle PDF effects from potential new physics in LHC data interpretations. Finally, neutrino measurements at the LHC will also test a variety of neutrino-philic BSM physics models, such as non-standard interactions~\cite{Ismail:2020yqc}, neutrino-electromagnetic properties~\cite{MammenAbraham:2023psg} and neutrino-philic new particles~\cite{Kling:2020iar,Kelly:2021mcd}.

\section*{Acknowledgments}

We acknowledge useful discussions with the FASER and SND@LHC Collaborations and the Forward Physics Facility coordination group. This work is supported by the Deutsche Forschungsgemeinschaft under Germany's Excellence Strategy -- EXC 2121 Quantum Universe -- 390833306. 


\end{document}